\documentclass[sigconf]{acmart}

\usepackage{graphicx}
\usepackage{epstopdf}

\usepackage{algorithm}
\usepackage{algorithmic}

\usepackage{array}
\newcommand{\PreserveBackslash}[1]{\let\temp=\\#1\let\\=\temp}
\newcolumntype{C}[1]{>{\PreserveBackslash\centering}p{#1}}
\newcolumntype{R}[1]{>{\PreserveBackslash\raggedleft}p{#1}}
\newcolumntype{L}[1]{>{\PreserveBackslash\raggedright}p{#1}}

\setcopyright{none}

\acmDOI{00.000/000_0}

\acmISBN{000-0000-00-000/00/00}

\acmConference[0000]{000}{0000}{Beijing, China} 
\acmYear{0000}
\copyrightyear{0000}

\acmPrice{00.00}

\begin{document}
	\title{Understanding Graph and Understanding Map and their Potential Applications}
	
	\author{Gangli Liu}
	\affiliation{%
		\institution{Tsinghua University}
		\city{Beijing} 
		\state{China} 
		\postcode{100084}
	}
	\email{gl-liu13@mails.tsinghua.edu.cn}

	\begin{abstract}
		Based on the previously proposed concept Understanding Tree, this paper introduces two concepts: Understanding Graph and Understanding Map, and explores their potential applications. Understanding Graph and Understanding Map can be deemed as special cases of mind map, semantic network, or concept map. The two main differences are: Firstly, the data sources for constructing Understanding Map and Understanding Graph are distinctive and simple. Secondly, the relations between concepts in Understanding Graph and Understanding Map are monotonous.  Based on their characteristics, applications of them include quantitatively measuring a concept's complexity degree, quantitatively measuring a concept's importance degree in a domain, and computing an optimized learning sequence for comprehending a concept etc. Further study involves evaluating their performances in these applications. 
	\end{abstract}
	
	%
	%

	

	\maketitle
	
	\section{Introduction}
	
	In \cite{Liu2016a}, Understanding Tree is introduced for evaluating a person's understanding degree to a piece of knowledge. This paper introduces two more concepts, Understanding Graph and Understanding Map, and explores their potential applications. We first recap several concepts defined in previous papers \cite{Liu2016,Liu2016a}. Section 2 tells about Understanding Graph and its applications. Section 3 gives an account of Understanding Map and its applications. Section 4 compares Understanding Graph and Understanding Map with other knowledge organizing and representing tools. Section 5 concludes the paper.
	\subsection{Recap of several definitions}
	Here we recap several concepts mentioned in previous papers.
	\subsubsection*{Knowledge Point}
	A Knowledge Point is a piece of knowledge which is explicitly defined and has been widely accepted. A Knowledge Point's definition may be differently phrased, however, they should be consistent.
	
	\subsubsection*{Basic Knowledge Point (BKP)}
	
	A BKP is a Knowledge Point that is simple enough so that it is not interpreted by other Knowledge Points.  In a sense, they are like axioms in mathematics, serving as premises or starting points for further reasoning and arguments. In practice, it is subjective which set of Knowledge Points should be categorized as BKPs. It can be decided by a group of experts empirically.
	
	\subsubsection*{Familiarity Measure}
	A Familiarity Measure is a score that depicts a person's familiarity degree to a Knowledge Point at a particular time. It is calculated with the formulas mentioned in \cite{Liu2017}.
	
	\subsubsection*{Understanding Tree}
	An Understanding Tree is a treelike data structure which compiles the background Knowledge Points that are essential to understand the root Knowledge Point. 
	
	\subsubsection*{Fully Extended Understanding Tree (FEUT)}
	
	A fully extended Understanding Tree is a tree like Figure 3 of \cite{Liu2016a}. It can be constructed manually by some experts of a domain, or automatically with Algorithm 1 of \cite{Liu2016a}. Its construction is based on a collection of definitions that are self-contained. `Self-contained' means for every Knowledge Point mentioned in a definition, either it is a BKP, or it has at least one definition in the collection. The root of an FEUT is the Knowledge Point at which the construction starts.
	If every non-BKP has only one definition, it is intuitive to construct the FEUT. If a non-BKP has multiple definitions, and they are equivalent (e.g., the term `topological space' has several equivalent definitions), how to process the definitions is a problem. There are several strategies for dealing with it:
	\begin{itemize}
		\item Accept all its definitions and all the Knowledge Points mentioned in the definitions;
		\item Pick one definition and only accept the Knowledge Points mentioned in this definition;
		\item Use other rules like the one mentioned in Algorithm 1 of \cite{Liu2016a}.
	\end{itemize}
	
	\subsubsection*{Standard Understanding Tree (SUT)}
	A standard Understanding Tree is a tree that is generated from an FEUT, by removing identical nodes from it. 
	Figure 1 is an example of SUT. It is similar to Figure 4 of \cite{Liu2016a}, except that the BKPs are labeled with shading and the Knowledge Points are not tagged with Familiarity Measures. Figure 2 is another example of SUT. It is constructed according to the definitions of Table 1.
	
	\begin{figure}
		\centering
		\includegraphics[width=0.9\columnwidth]{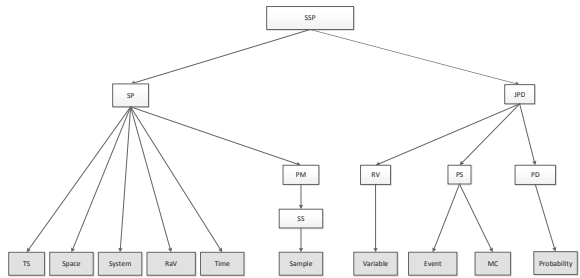}
		\caption{SUT of Strictly Stationary Process}
	\end{figure}
	
	\begin{figure}
		\centering
		\includegraphics[width=0.9\columnwidth]{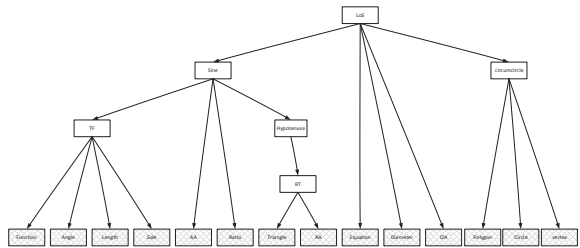}
		\caption{SUT of The Law of Sines}
	\end{figure}

	\begin{table*}
		\small
		\centering
		\begin{tabular}{|C{2cm}|L{11cm}|C{3.5cm}|}
			\hline
			Knowledge Point & \qquad \qquad \qquad \qquad \qquad\qquad \qquad Definition & Involved Knowledge Points  \\ \hline
			law of cosines (LoC) & The law of cosines relates the lengths of the sides of a triangle to the cosine of one of its angles. the law of cosines states:
			$ c^{2}=a^{2}+b^{2}-2ab\cos \gamma $,  
			where $ \gamma $ denotes the angle contained between sides of lengths $ a $ and $ b $ and opposite the side of length $ c $.  &  length, side, triangle, cosine, angle \\ \hline
			\qquad \qquad \qquad \qquad\qquad cosine &  The trigonometric function that is equal to the ratio of the side adjacent to an acute angle (in a right triangle) to the hypotenuse.  &  trigonometric function (TF) , ratio, side, acute angle (AA), right triangle (RT), hypotenuse \\ \hline
			Pythagorean theorem (PT) &  The Pythagorean theorem states that the square of the hypotenuse is equal to the sum of the squares of the other two sides.  &  square, hypotenuse, sum, side \\ \hline
			hypotenuse &  A hypotenuse is the longest side of a right triangle, the side opposite the right angle. &  side, right triangle (RT), right angle (RA) \\ \hline
			
			right triangle &  A triangle with a right angle.  &  triangle, right angle (RA) \\ \hline
			\qquad \qquad\qquad\qquad law of sines	(LoS)
			&  The law of sines is an equation relating the lengths of the sides of a triangle to the sines of its angles. According to the law,
			\begin{equation*}
				\frac{a}{\sin A} = \frac{b}{\sin B} = \frac{c}{\sin C} = d
			\end{equation*}
			where $ a, b $, and $ c $ are the lengths of the sides of a triangle, and $ A, B $, and $ C $ are the opposite angles, while $ d $ is the diameter of the triangle's circumcircle.  &  equation, length, side, triangle, sine, angle, opposite angle (OA), diameter, circumcircle \\ \hline
			\qquad\qquad\qquad\qquad\qquad sine & The sine is a trigonometric function of an angle. The sine of an acute angle is defined in the context of a right triangle: for the specified angle, it is the ratio of the length of the side that is opposite that angle to the length of the longest side of the triangle (the hypotenuse). &  trigonometric function (TF), angle, acute angle (AA), right triangle (RT), ratio, length, side, triangle, hypotenuse \\ \hline
			circumcircle & The circumcircle of a polygon is a circle which passes through all the vertices of the polygon. &  polygon, circle, vertex \\ \hline 	 
			
			trigonometric function (TF) & The trigonometric functions are functions of an angle. They relate the angles of a triangle to the lengths of its sides. &  function, angle, triangle, length, side \\ 
			\hline 
		\end{tabular}
		\caption{A set of self-contained definitions}
	\end{table*}

	\section{Understanding Graph}
	An Understanding Graph is a rooted graph that is also generated from an FEUT. Instead of removing identical nodes from an FEUT, it merges them.  The merging of identical nodes keeps the connection relations (links and their directions) between Knowledge Points unchanged; only the nodes that representing the same Knowledge Point are merged into one node.
	E.g., Figure 3 is an Understanding Graph constructed based on Figure 3 of \cite{Liu2016a}. Figure 4 is another Understanding Graph constructed based on the definitions of Table 2. The definitions listed in Table 1 and 2 are cited from Wikipedia and other authoritative websites like Wolfram MathWorld. The shaded nodes in Figure 3 and 4 are BKPs. In Understanding Graph, each Knowledge Point is represented by one and only one node. If two Knowledge Points have the same name but with different meanings (such as homonyms), or a concept has a particular meaning in a specific field (such as the term `compact' in Topology), they are differentiated with different nodes. If a Knowledge Point has multiple names, such as the law of sines, they are incorporated into one node. 
	
	\begin{figure}
		\centering
		\includegraphics[width=0.9\columnwidth]{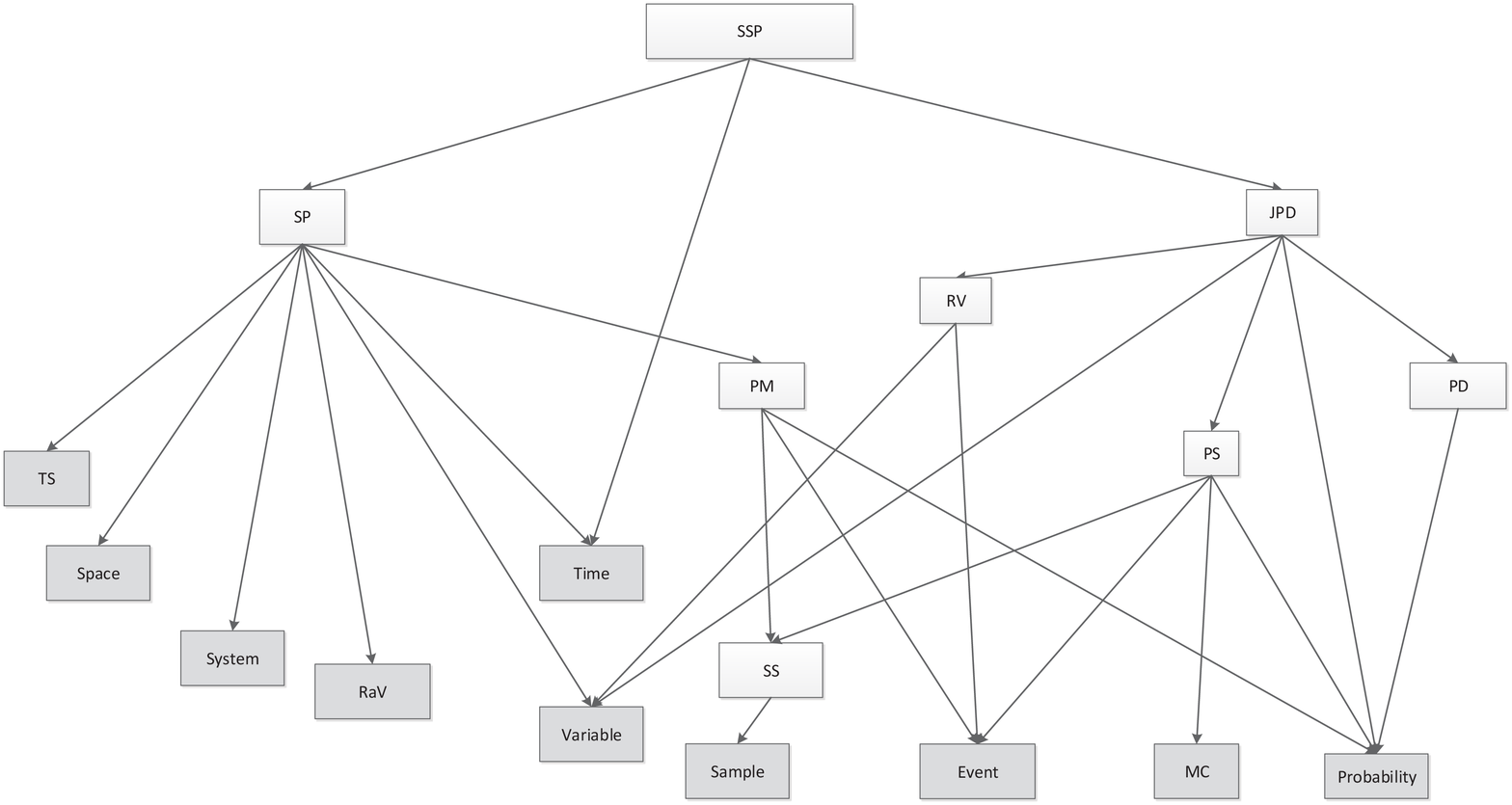}
		\caption{Understanding Graph of Strictly Stationary Process}
	\end{figure}
	
	\begin{figure*}
		\centering
		\includegraphics[width=1.9\columnwidth]{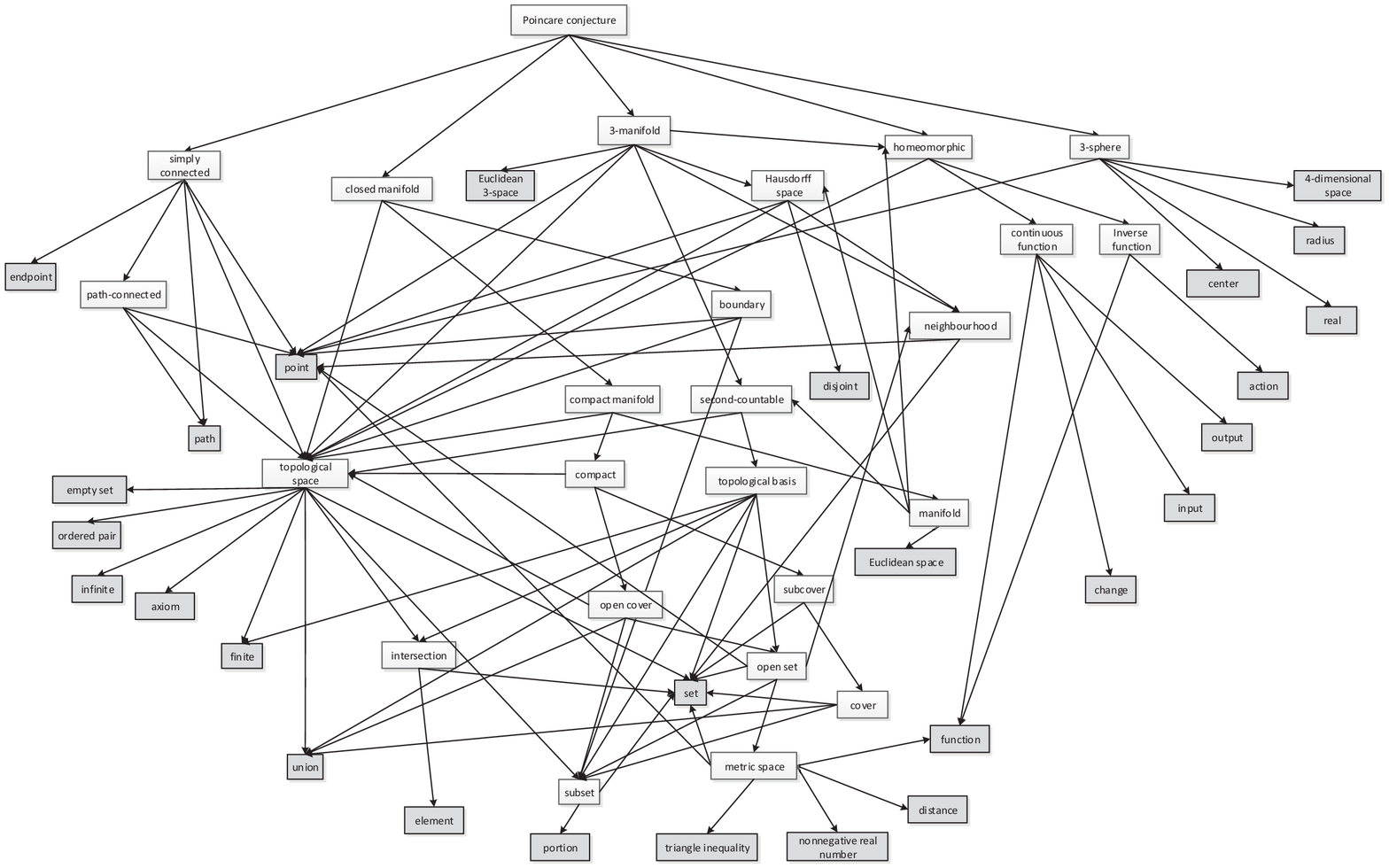}
		\caption{Understanding Graph of the Poincare Conjecture}
	\end{figure*}
	
	\begin{table*}
		\small
		\centering
		\begin{tabular}{|C{2cm}|L{10cm}|C{4.5cm}|}
			\hline
			Knowledge Point & \qquad \qquad \qquad \qquad \qquad\qquad \qquad Definition & Involved Knowledge Points  \\ \hline
			Poincare conjecture & Every simply connected, closed 3-manifold is homeomorphic to the 3-sphere.  &  simply connected,	closed,	3-manifold,	homeomorphic, 3-sphere
			\\ \hline
			simply connected &  A topological space is called simply-connected if it is path-connected and every path between two points can be continuously transformed, staying within the space, into any other such path while preserving the two endpoints in question.  &  topological space, path-connected,
			path, point, endpoint \\ \hline
			closed manifold &  A closed manifold is a type of topological space, namely a compact manifold without boundary.   &  topological space, compact manifold, boundary \\ \hline
			\qquad\qquad\qquad\qquad3-manifold &  A topological space $ X $ is a 3-manifold if it is a second-countable Hausdorff space and if every point in $ X $ has a neighborhood that is homeomorphic to Euclidean 3-space. &  topological space, second-countable, Hausdorff space,	point, neighborhood, homeomorphic,	Euclidean 3-space \\ \hline		
			homeomorphic &  A homeomorphism is a continuous function between topological spaces that has a continuous inverse function. &  continuous function, topological space, inverse function \\ \hline
			\qquad \qquad\qquad\qquad the 3-sphere
			&  A 3-sphere with center $ (C0, C1, C2, C3) $ and radius $ r $ is the set of all points $ (x0, x1, x2, x3) $ in real, 4-dimensional space $ (R4) $ such that
			\begin{equation*}
				\sum_{i=0}^3{(x_{i} - C_{i})}^2 = {(x_{0} - C_{0})}^2 + {(x_{1} - C_{1})}^2 + {(x_{2} - C_{2})}^2 + {(x_{3} - C_{3})}^2 = r^2
			\end{equation*}
			&  center, radius, point, real, 4-dimensional space  \\ \hline
			\qquad\qquad\qquad topological space & A topological space is an ordered pair $ (X, \tau) $, where $ X $ is a set and $ \tau $ is a collection of subsets of $ X $, satisfying the following axioms: (1) The empty set and $ X $ itself belong to $ \tau $ . (2)	Any (finite or infinite) union of members of $ \tau $ still belongs to $ \tau $. (3) The intersection of any finite number of members of $ \tau $ still belongs to $ \tau $. &  ordered pair,	set, subset, axiom,	empty set, finite, infinite, union, intersection \\ \hline
			path-connected & A topological space is said to be path-connected if given any two points on the topological space, there is a path starting at one point and ending at the other. &  topological space, point, path \\ \hline 	 		
			compact manifold & A compact manifold is a manifold that is compact as a topological space. &  manifold, compact, topological space \\ \hline 
			boundary	&The boundary of a subset $ S $ of a topological space $ X $ is the set of points which can be approached both from $ S $ and from the outside of $ S $.	&subset, topological space, point	\\ \hline
			Hausdorff space	&The Hausdorff space is a topological space in which distinct points have disjoint neighborhoods.	&topological space, point, disjoint, neighborhood	\\ \hline
			second-countable	&A topological space is second countable if it has a countable topological basis.	&topological space, topological basis	\\ \hline
			neighborhood	&A neighborhood of a point is a set of points containing that point where one can move some amount away from that point without leaving the set.	&Point, set	\\ \hline
			continuous function	&A continuous function is a function for which sufficiently small changes in the input result in arbitrarily small changes in the output.	&function, change, input, output	\\ \hline
			inverse function	&An inverse function is a function that undoes the action of another function.	&function, action	\\ \hline
			manifold	&A manifold is a second countable Hausdorff space that is locally homeomorphic to Euclidean space.	&second countable, Hausdorff space, homeomorphic, Euclidean space	\\ \hline
			compact	&A topological space is compact if every open cover of $ X $ has a finite subcover.	&topological space, open cover, subcover	\\ \hline
			topological basis&	A topological basis is a subset $ B $ of a set $ T $ in which all other open sets can be written as unions or finite intersections of $ B $.&	subset, set, open set, union, finite, intersection \\ \hline
			open cover&	A collection of open sets of a topological space whose union contains a given subset.	&open set, topological space, union, subset\\ \hline
			subcover&	Let $ S $ be a set.	Let $ U $ be a cover for $ S $.	A subcover of $ U $ for $ S $ is a set $ V \subseteq U $ such that $ V $ is also a cover for $ S $. &	set, cover\\ \hline
			open set&	Let $ S $ be a subset of a metric space. Then the set $ S $ is open if every point in $ S $ has a neighborhood lying in the set.&	subset, metric space, set, point, neighborhood\\ \hline
			intersection&	The intersection of two sets $ A $ and $ B $ is the set of elements common to $ A $ and $ B $.&	set, element\\ \hline
			subset&	A subset is a portion of a set.&	portion, set\\ \hline
			cover&	A cover of a set $ X $ is a collection of sets whose union contains $ X $ as a subset.&	set, union, subset\\ \hline
			\qquad\qquad\qquad\qquad metric space&	A metric space is a set $ S $ with a global distance function that, for every two points $ x,y $ in $ S $, gives the distance between them as a nonnegative real number $ g(x,y) $. A metric space must also satisfy	(1) $ g(x,y)=0$ iff $x=y $, (2) $ g(x,y)=g(y,x) $, (3) The triangle inequality $ g(x,y)+g(y,z)>=g(x,z) $. &	set, function, point, distance, nonnegative real number, triangle inequality\\ \hline		
		\end{tabular}
		\caption{Another set of self-contained definitions}
	\end{table*}
	
	\subsection{Application of Understanding Graph}
	Several potential applications have been devised for Understanding Graph.
	
	\subsubsection{Extending keywords in Information retrieval}	
	Information retrieval (IR) is the activity of obtaining information resources relevant to an information need from a collection of information resources. When a person has an information need, he inputs some keywords to a search engine. Usually, the more keywords are submitted, the more a search engine can know the user's information need. However, more keywords require more cognitive burden for the user to formulate and input. With Understanding Graph, we can extend the user's keywords automatically. E.g., if the user inputs \emph{`Poincare Conjecture'}, it seems the user wants to know what Poincare Conjecture is, then the keyword is extended into \emph{``Poincare Conjecture, simply connected, closed manifold, homeomorphic, topological space, Hausdorff space, neighborhood ...''} The extension can be accomplished according to a Knowledge Point's SUT, from low level nodes to high level nodes. A document or a series of documents containing all the extended keywords are good candidates for the information need.
	
	\subsubsection{Indicating the minimum  information for understanding a Knowledge Point}
	
	Because an Understanding Graph is just a transformation of an Understanding Tree. The Understanding Tree of a Knowledge Point is constructed based on the definitions of itself and its descendants. It compiles the most essential information for understanding a Knowledge Point. Knowing other Knowledge Points may help understand it, but they are not the most necessary ones. E.g., knowing the Pythagorean Theorem is a special case of the Law of Cosines (LoC) helps understand the LoC, but it is not essential for understanding it. Thus it indicates the minimum information for understanding a Knowledge Point. 
	
	\subsubsection{Indicating an optimized learning sequence for comprehending a Knowledge Point}
	
	In meaningful learning, the learners are `integrating' new information into old information \cite{novak2002meaningful}. Meaningful learning is opposed to rote learning and refers to a learning method where the new knowledge to acquire is related with previous knowledge \cite{ausubel2012acquisition}. If a person is facing a piece of information full of concepts he does not understand, he cannot practice a meaningful learning. E.g., if you tell a person who knows little about Mathematics the content of Poincare Conjecture (now it is a theorem):\emph{``Every simply connected, closed 3-manifold is homeomorphic to the 3-sphere.''}	
	He will be perplexed by what a manifold is and what do you mean by saying \emph{`simply connected', `closed',} and \emph{`homeomorphic'.} To practice a meaningful learning, a person should have understood these concepts before facing this information. Understanding Graph provides a way to comprehend a complicated Knowledge Point in a from-easy-to-difficult order. 
	
	Take the Poincare Conjecture as an example, Figure 4 is its Understanding Graph. To learning it meaningfully, a person can lean the BKPs first. After comprehended them, they are removed from the Understanding Graph. The graph becomes Figure 5. The next learning target is determined by checking the out-degrees of the nodes of Figure 5. The nodes with the smallest out-degrees are candidates (the light blue ones). When a node is learned and understood, it is removed from the Understanding Graph, then compares the out-degrees of the rest nodes to determine the next target. The learning process ends until the last node is removed. Because of the decaying of human memory, what have been learned may be forgotten, a person can review previously learned information at any time during the learning process. 
	
	\begin{figure*}
		\centering
		\includegraphics[width=1.7\columnwidth]{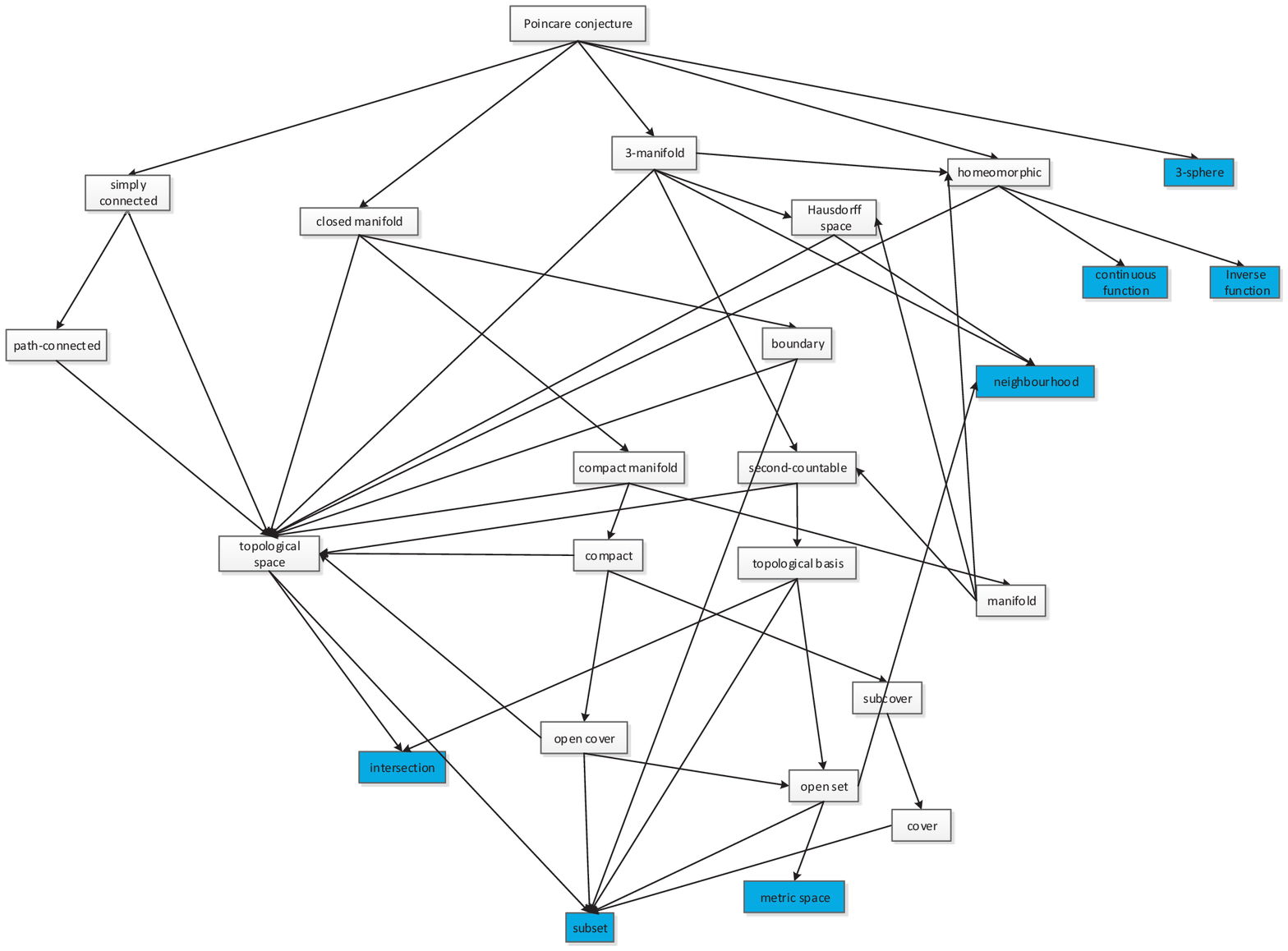}
		\caption{Understanding Graph of the Poincare Conjecture without BKPs}
	\end{figure*}
	
	Algorithm 1 calculates an optimized learning sequence for comprehending a Knowledge Point, based on its Understanding Graph. The subject is supposed having understood the BKPs before learning, so the learning sequence does not consider BKPs.
	
	\begin{algorithm}
		\caption{Calculating an optimized learning sequence for comprehending a Knowledge Point based on its Understanding Graph}  
		\begin{algorithmic}[1] 
			\REQUIRE ~~\\ 
			Knowledge Point $ k_{i} $'s Understanding Graph, $ G_{i} $;		
			\ENSURE ~~\\ 
			An optimized learning sequence for comprehending $ k_{i} $, $ S $;
			\STATE Set $ S = NULL $;
			\STATE Remove the BKPs and their edges from $ G_{i} $;
			\STATE Check whether $ G_{i} $ is $ NULL $. If so, go to step 6;
			\STATE Check the out-degrees of the nodes of $ G_{i} $. Select the node with the smallest out-degree, record it as $ n_{i} $ (if there are more than one such nodes, select one arbitrarily). 
			\STATE Insert $ n_{i} $ into $ S $. Remove $ n_{i} $ and its edges (to or from $ n_{i} $) from $ G_{i} $. Go to step 3;
			\RETURN $ S $.			
		\end{algorithmic}
	\end{algorithm}
	
	According to Algorithm 1 and Understanding Graph Figure 4, one of the suggested learning sequences for comprehending Poincare Conjecture is: \emph{intersection, subset, metric space, neighborhood, continuous function, inverse function, 3-sphere, open set, cover, subcover, topological space, topological basis, open cover, compact, path-connected, simply connected, homeomorphic, boundary, Hausdorff space, second-countable, manifold, compact manifold, closed manifold, 3-manifold, Poincare Conjecture.}
	
	Algorithm 3 of \cite{Liu2016a} also facilitates meaningful learning. The difference between Algorithm 1 and Algorithm 3 of \cite{Liu2016a} is: for Algorithm 3 of \cite{Liu2016a}, there are a collection of documents to be learned, the goal is to learn the documents meaningfully. The algorithm checks the subject's current knowledge status and suggests the most appropriate document for current learning; for Algorithm 1, there is no corpus to be learned. The goal is to understand the root Knowledge Point by meaningful learning. The algorithm outputs a plan for achieving the goal. The learning process is divided into small steps. In each step, there is a small target to learn and comprehend an intermediate Knowledge Point that is essential for understanding the root. The subject is free to choose any methods to accomplish this small target, such as reading a book, surfing the Internet, or discussing with others etc. The only requirement is that the subject has understood the intermediate Knowledge Point after that step.
	The sequence is `optimized' in that it recommends the easier ones first, then the harder ones, and comprehending the harder ones relies on understanding the easier ones. In a sense, the learning process is similar to let a group of sappers clear a minefield first, before troops pass through the field. An Understanding Graph is like a map tagging where the mines are located and how they are connected; Algorithm 1 calculates an optimized sequence for clearing the mines. Therefore, it facilitates a meaningful learning path for comprehending the root Knowledge Point.
	
	\subsubsection{Measuring a Knowledge Point's complexity degree}
	
	In Understanding Graph, a Knowledge Point is represented as a bag-of-concepts and their relations. By examining the size of a Knowledge Point's Understanding Graph, its complexity degree can be obtained. E.g., by estimating empirically, we know that Poincare Conjecture is more complicated than Pythagorean Theorem. Understanding Graph facilitates a quantitative description of this relation. A Knowledge Point's complexity degree is defined as the number of vertices plus the number of edges of its Understanding Graph. Hence according to Poincare Conjecture and Pythagorean Theorem's Understanding Graphs (Figure 4 and Figure 6), the complexity degree of Poincare Conjecture equals 148; the complexity degree of Pythagorean Theorem equals 17. Based on this definition, Poincare Conjecture is indeed more complicated than Pythagorean Theorem.
	
	\begin{figure}
		\centering
		\includegraphics[width=0.6\columnwidth]{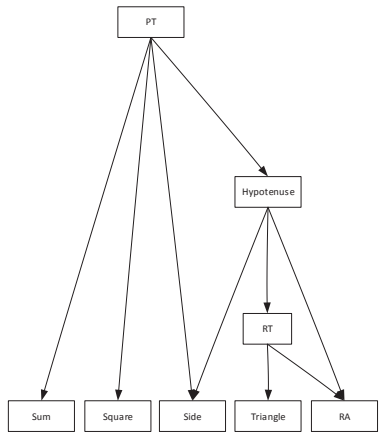}
		\caption{Understanding Graph of the Pythagorean Theorem}
	\end{figure}
	
	In \cite{Liu2016a}, I propose using the height and number of nodes of a Knowledge Point's SUT to characterize its complexity degree. In this article, I propose another method for measuring its complexity degree.
	\section{Understanding Map}
	Suppose we have constructed Understanding Graphs for a group of Knowledge Points, a map can be generated by merging the graphs. Understanding Map is defined as a graph which is constructed by merging a set of Understanding Graphs (at least one). The merging combines identical nodes into one node and keeps the link relations unchanged.
	E.g., Figure 7 is an Understanding Map constructed based on four Understanding Graphs:  LoC, PT, LoS, and SSP. The Understanding Graphs of LoC, PT, and LoS are constructed based on the definitions of Table 1. Figure 3 is the Understanding Graph of SSP. Figure 4 is the Understanding Graph of Poincare Conjecture. It can also be deemed as an Understanding Map. Understanding Graph and Understanding Map are devised for different applications. The main difference is that an Understanding Graph is at least weakly connected; an Understanding Map does not need to be connected, like the one in Figure 7. Understanding Map usually contains large scale of Knowledge Points. If we combine all Knowledge Points' Understanding Graphs, a very large Understanding Map is generated. It is called the Global Understanding Map (GUM), like a world map.
	
	\begin{figure*}
		\centering
		\includegraphics[width=1.3\columnwidth]{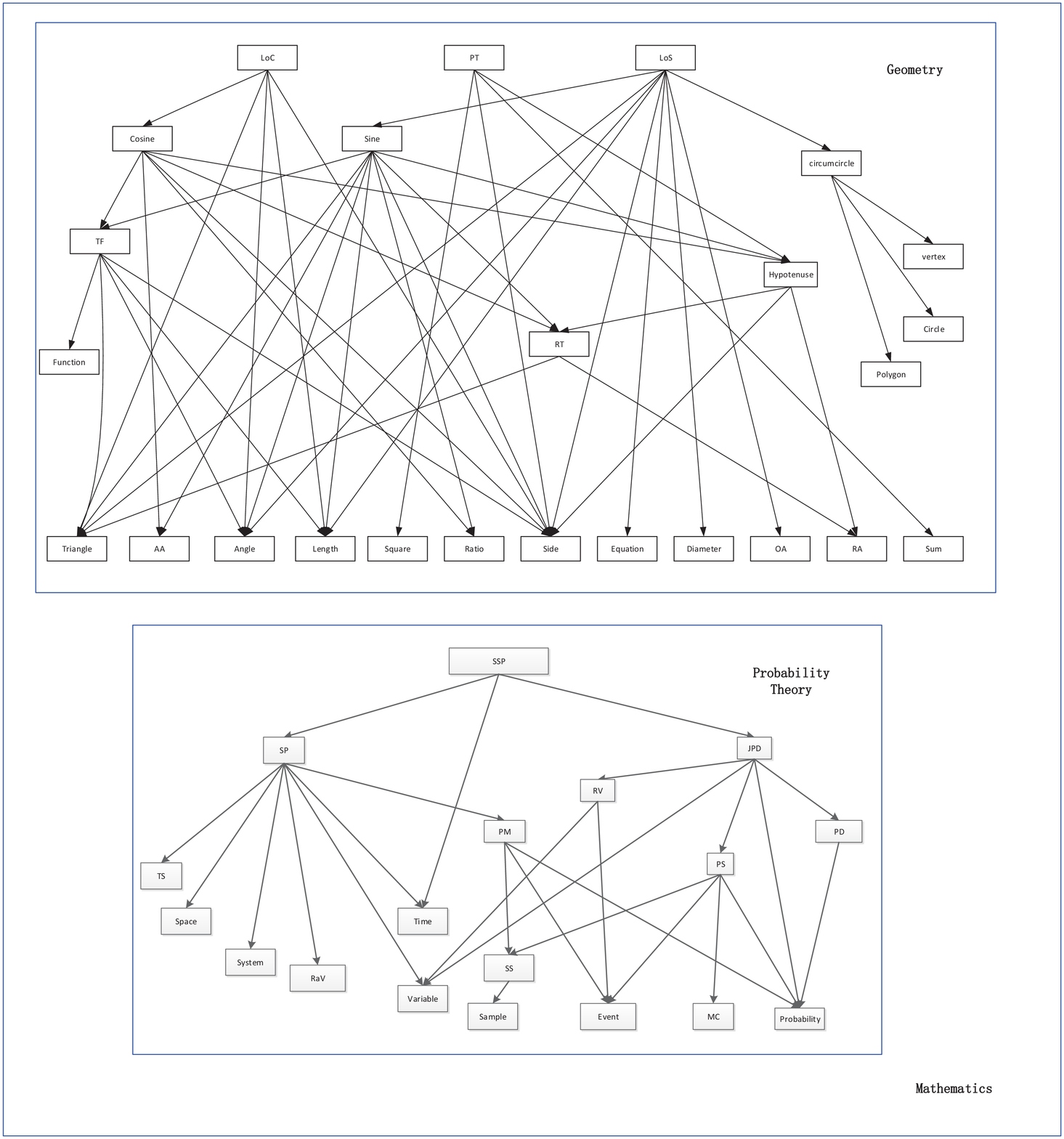}
		\caption{An exemplary Understanding Map}
	\end{figure*}
	
	\subsection{A node's n-level neighborhood}
	
	A node's n-level open neighborhood in an Understanding Map is defined as a set of nodes whose undirected distance to the node is less than or equal to $ n $, excluding the node itself. The n-level closed neighborhood is defined in the same way but also includes the node itself. The n-level neighborhood is an extension of the term `neighborhood' in graph theory. Before counting distances, the edges in Understanding Map are converted to be undirected. The distance between any two vertices in a graph is the length of the shortest path having the two vertices as its endpoints.  E.g., in the Understanding Map of Figure 8, the 1-level open neighborhood of the node `topological basis' is a set that is comprised of the green nodes; the 2-level open neighborhood is comprised of the green nodes plus the yellow nodes.
	
	\begin{figure*}
		\centering
		\includegraphics[width=1.9\columnwidth]{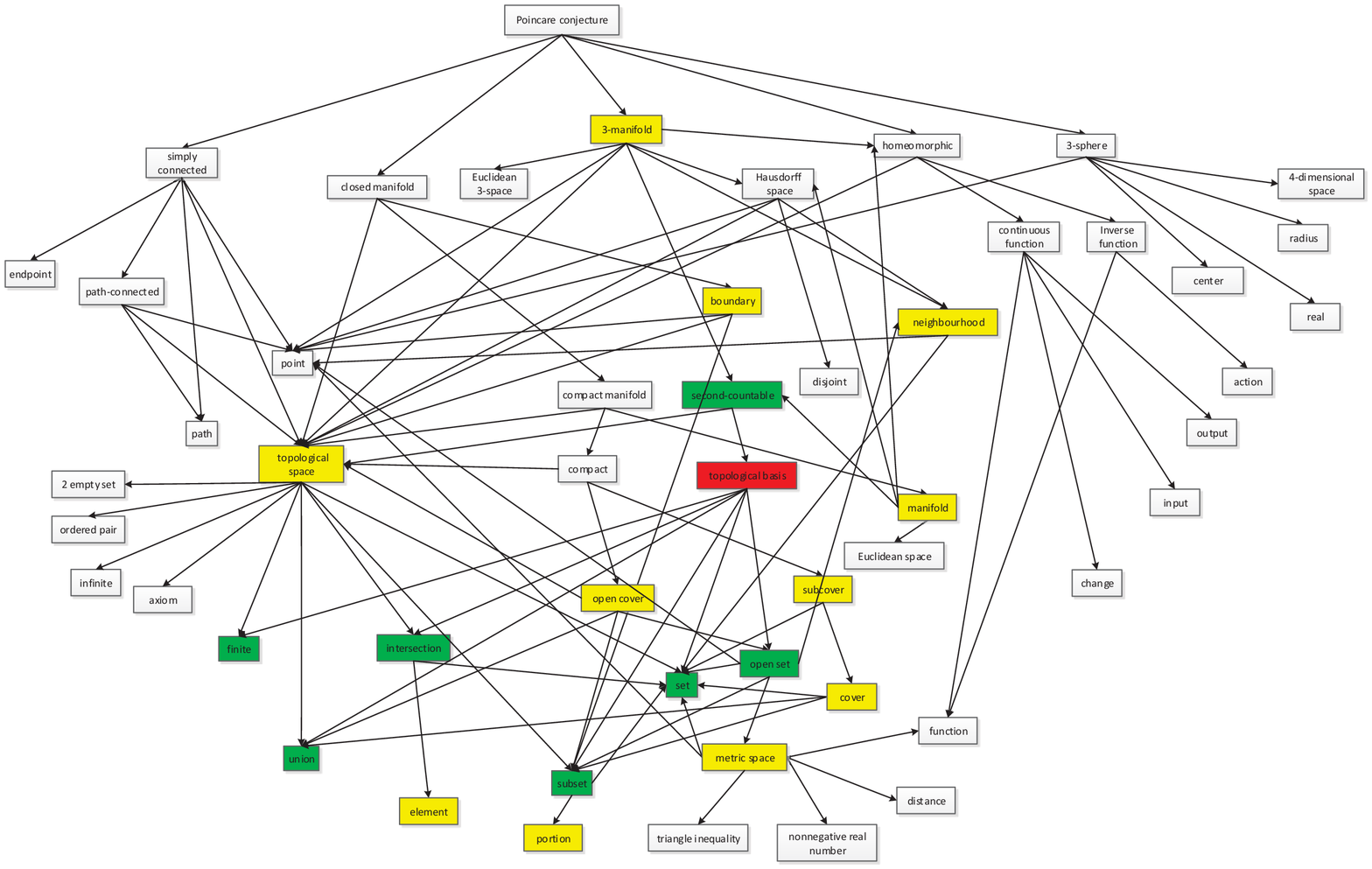}
		\caption{A node's n-level neighborhood in an Understanding Map}
	\end{figure*}
	
	\subsection{Application of Understanding Map}
	The following are some potential applications of Understanding Map.
	
	\subsubsection{Differentiating Homograph Knowledge Points (HKP)}
	
	A HKP is defined as a Knowledge Point that shares the same written form with another Knowledge Point but has a different meaning. It is an extension of the term `homograph' in linguistics.  E.g., the Knowledge Point `scale' is an HKP; it has different meanings in different context, such as:
	\begin{itemize}
		\item Scale (anatomy), a small rigid plate that grows out of an animal's skin to provide protection.
		\item Scale (map), the ratio of the distance on a map to the corresponding actual distance.
		\item Scale (measuring), an instrument used to measure weight.
		\item Scale (music), a set of musical notes ordered by fundamental frequency or pitch.
	\end{itemize}
	In practice, people usually can differentiate a HKP's meaning by its context without much effort, but it is not easy for a machine. Algorithm 2 provides a method for differentiating an HKP based on an Understanding Map $ U $ (preferably the Global Understanding Map) and a context $ C $ in which the HKP appears. Context $ C $ is represented by a small bag-of-words, coming from one or more sentences, such as the sentences that containing or adjoining the HKP. In Understanding Map $ U $, each meaning of the HKP possesses a different node; it is called a concrete of the HKP. Understanding Map $ U $ contains all Understanding Graphs that are related to any one of the concretes. Each concrete of the HKP has a different neighborhood in Understanding Map $ U $. 
	Algorithm 2 works by matching each concrete's n-level open neighborhood with the HKP's current context $ C $. It starts at $ n $ equals 1. A concrete wins when it matches the most Knowledge Points with context $ C $. If no concrete wins when $ n $ equals 1, then increase $ n $ by one level and compares again, until $ n $ equals $ k $. If the winner still cannot be decided at last. The algorithm returns a $ NULL $, indicating it has failed for differentiating the HKP.
	
	\begin{algorithm}
		\caption{An algorithm for differentiating HKP}  
		\begin{algorithmic}[1] 
			\REQUIRE ~~\\ 
			Understanding Map, $ U $;
			Context, $ C $;
			Homograph Knowledge Point, $ H $;
			Maximum attempts, $ k $;	
			\ENSURE ~~\\ 
			One concrete of $ H $ according to context $ C $, $ h_{i} $;
			
			\STATE Set $ n = 1$, $ h_{i} = NULL $;
			\STATE If $ n > k $, go to step 6;
			\STATE Calculate each concrete's n-level open neighborhood;
			\STATE Match the n-level open neighborhoods with context $ C $. If there is only one concrete matching the most Knowledge Points, set $ h_{i} $ equals it, go to step 6;
			\STATE Increase $ n $ by one level. Go to step 2;
			\RETURN $ h_{i} $.			
		\end{algorithmic}
	\end{algorithm}
	
	\subsubsection{Another way for calculating understanding degree}
	
	In \cite{Liu2016a}, I have devised a method for calculating a person's understanding degree to a Knowledge Point, by examining his Familiarity Measures to the Knowledge Point's SUT. This article proposes another way for calculating the understanding degree, based on an Understanding Map (preferably the GUM). It calculates the subject's average Familiarity Measure to a Knowledge Point's n-level closed neighborhood (n is a parameter to be determined), then uses the average value as the understanding degree.
	
	\subsubsection{Illustrating the knowledge characteristics of a person or a corpus}
	
	An Understanding Map is analogous to a real map. E.g., in a sense, Figure 7 is like Figure 9. A Knowledge Point is like a building; closely related Knowledge Points constitute a district, a city, or a country, just like a domain or discipline. A person lives in a city, he must be familiar with the buildings of the city than other cities. It is analogous to a person having expertise in a domain, and being familiar with the topics of the domain. In \cite{Liu2016} and \cite{Liu2017} , I propose a formula for calculating a person's familiarity degree to a Knowledge Point at a particular time. If we display the familiarity degrees in an Understanding Map, by setting a node's brightness being proportional to a person's familiarity degree to the node, a night light map like Figure 10 can be obtained. It illustrates a person's knowledge characteristics. A corpus can be illustrated similarly in an Understanding Map. A node's brightness can be set being proportional to its term frequency in the corpus.

	\begin{figure}
		\centering
		\includegraphics[width=0.7\columnwidth]{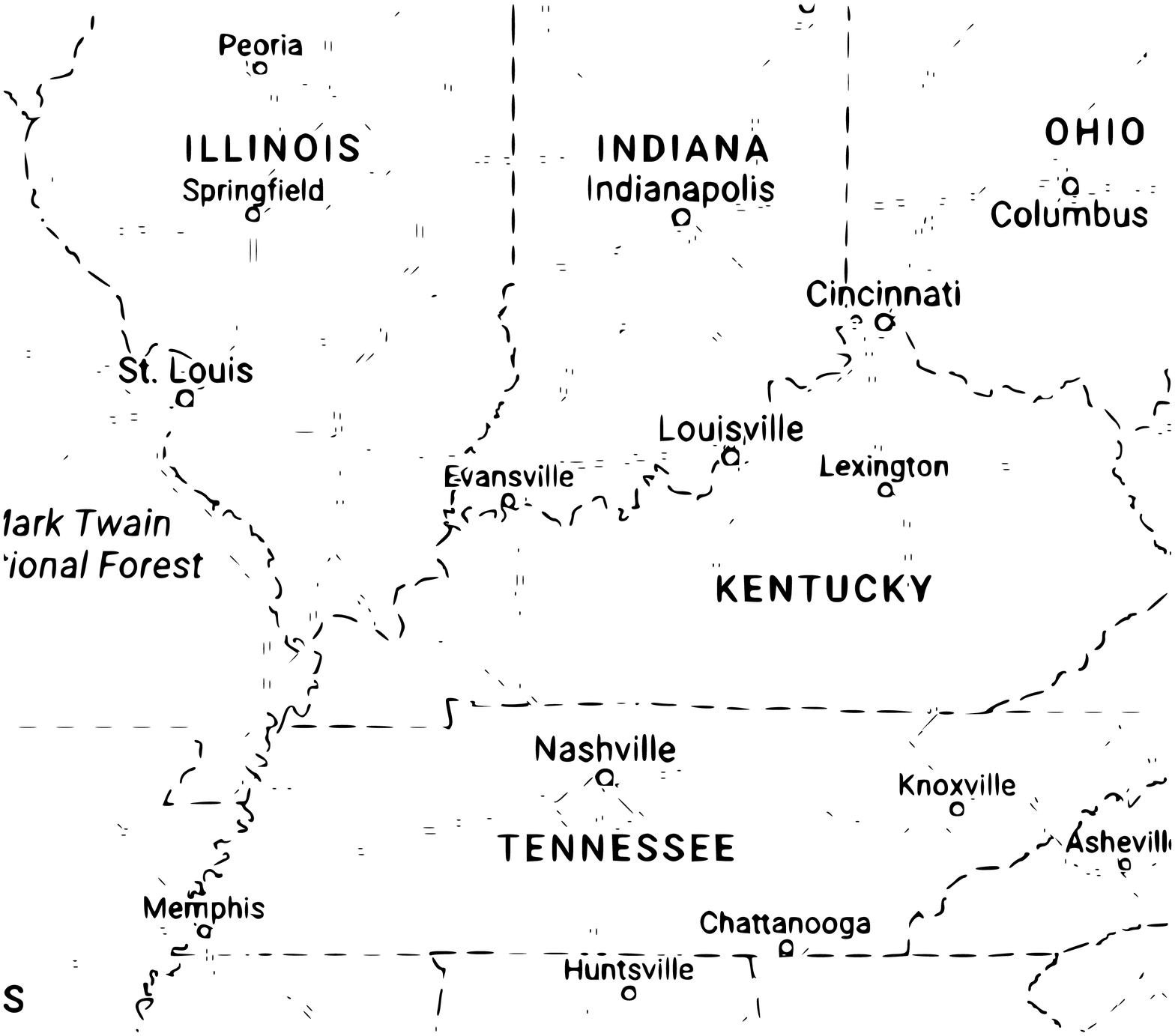}
		\caption{A part of a real map}
	\end{figure}
	
	\begin{figure}
		\centering
		\includegraphics[width=0.7\columnwidth]{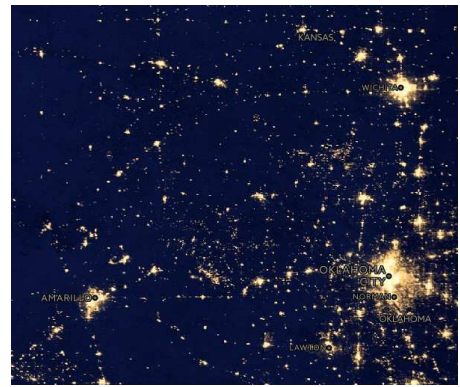}
		\caption{A part of a night light map}
	\end{figure}
	
	\subsubsection{Measuring a Knowledge Point's importance degree in a domain}
	
	In Understanding Graph and Understanding Map, an edge from node A to node B means understanding A relies on understanding of B. If a node is relied on by a lot of nodes, it implies the node is very important for understanding other nodes. Therefore, if we have an Understanding Map that contains all the Knowledge Points in a domain, by counting a node's in-degree, its importance rank in the domain is obtained. E.g., suppose we have a domain called ``Poincare Conjecture's definition'', also suppose the Understanding Map of Figure 4 contains all the Knowledge Points of this domain, by counting each node's in-degree, the top 3 important Knowledge Points in this domain are: topological space (11), point (9), and set (9). If we ignore the BKPs when ranking importance, the top 4 are: topological space (11), subset (6), homeomorphic (3), and neighborhood (3).
	
	\subsection{Conversion between Understanding Map, Understanding Graph, FEUT, and SUT}
	
	I have discussed how to convert an FEUT into an Understanding Graph and a set of Understanding Graphs into an Understanding Map. To obtain an Understanding Graph from an Understanding Map, firstly we designate the root node, then exclude nodes that are not descendants of the root.  Algorithm 3 converts an Understanding Graph into a SUT. Firstly, it decides each node's level according to the root, then removes redundant edges according to some rules. An FEUT can be converted into a SUT by firstly converting it into an Understanding Graph.
	
	\begin{algorithm}
		\caption{Converting an Understanding Graph into a SUT}  
		\begin{algorithmic}[1] 
			\REQUIRE ~~\\ 
			Understanding Graph, $ G $; The root Knowledge Point, $ R $;	
			\ENSURE ~~\\ 
			A SUT corresponding to G, $ S $;
			
			\STATE Decide each node's level according to $ R $. A node's level is the length of the shortest path from $ R $ to it (the path is directed); 
			\STATE Remove all the edges from higher level nodes to lower level ones, or between the same level from $ G $;
			\STATE Check each node's in-degree. If it is larger than one, keep one arbitrarily, remove others from $ G $;
			\STATE Set $ S = G $;
			\RETURN $ S $.			
		\end{algorithmic}
	\end{algorithm}

	Figure 11 is a SUT converted from the Understanding Graph of Figure 4. The nodes' levels are differentiated by different colors. In SUT, a node's level indicates how important it is for understanding the root.
	\begin{figure*}
		\centering
		\includegraphics[width=1.9\columnwidth]{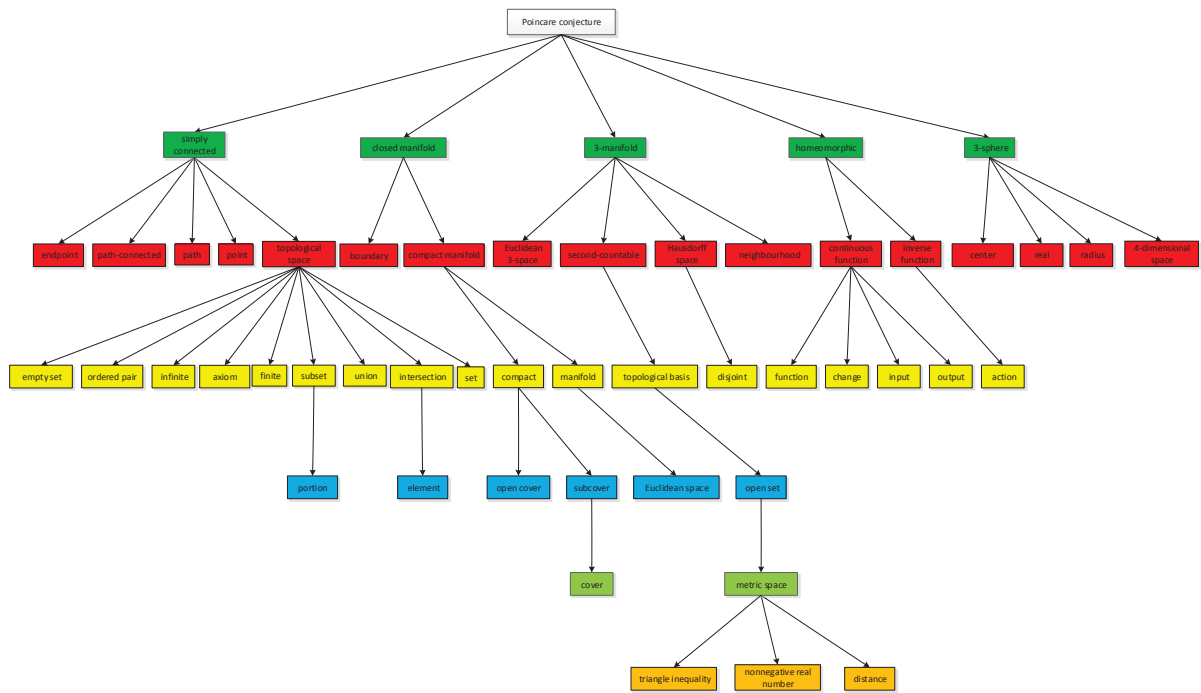}
		\caption{SUT of the Poincare Conjecture}
	\end{figure*}
	
	\section{Related work}
	A variety of tools have been devised for organizing concepts and their relationships, such as mind map, spider diagram, semantic network, and concept map etc.
	A mind map is a diagram used to visually organize information. It is hierarchical and shows relationships among pieces of the whole \cite{hopper2012practicing}. Mind maps are considered to be a type of spider diagram. A spider diagram is a boolean expression involving unitary spider diagrams and the logical symbols conjunction, disjunction, and negation. A unitary spider diagram adds existential points to an Euler or a Venn diagram \cite{howse2005spider}.
	A semantic network is a network that represents semantic relations between concepts. It is a directed or undirected graph consisting of vertices, which represent concepts, and edges, which represent semantic relations between concepts \cite{sowa2006semantic}. Typical standardized semantic networks are expressed as semantic triples. A semantic triple is a set of three entities that codifies a statement about semantic data in the form of subject-predicate-object expressions (e.g. ``a bear is a mammal'', or ``fish lives in water'').
	A concept map is a diagram that depicts suggested relationships between concepts \cite{hager1997designing}. It typically represents ideas and information as boxes or circles, which it connects with labeled arrows in a downward-branching hierarchical structure. The relationship between concepts can be articulated in linking phrases such as causes, requires, or contributes to \cite{novak2008theory}.
	
	Understanding Graph and Understanding Map can be considered as special cases of the tools listed above. The two main differences are: Firstly, the data sources for constructing Understanding Maps and Understanding Graphs are distinctive and simple. The sufficient and prerequisite condition for constructing an Understanding Graph is a set of self-contained definitions of Knowledge Points. Merging a set of Understanding Graphs (at least one) produces an Understanding Map. Secondly, the relations between concepts in Understanding Graph and Understanding Map are monotonous.  An edge from node A to node B always means understanding A relies on understanding of B. Since the relation is simplex, it is omitted in the graphs.

	\section{Conclusion}
	There should be a day, people will realize we have almost completed the knowledge edifice; it is time to organize our knowledge seriously, giving each concept a standardized, authoritative, and rigorous definition. Understanding Graph and Understanding Map provide a way for organizing all the well justified concepts, exploring their relationships according to their definitions. Besides proposing notions, I have explored several potential applications of the two data structures, such as quantitatively measuring a concept's complexity degree according to its Understanding Graph, quantitatively measuring a concept's importance degree in a domain according to the domain's Understanding Map, and computing an optimized learning sequence for comprehending a concept based on its Understanding Graph etc. Further study is necessary for evaluating their performances in these applications. In addition, three algorithms are devised. Algorithm 1 calculates the optimized learning sequence for comprehending a concept; Algorithm 2 differentiates an HKP based on an Understanding Map and a context in which the HKP appears; Algorithm 3 converts an Understanding Graph into a SUT.

	\bibliographystyle{ACM-Reference-Format}
	\bibliography{utree-umap} 
	
\end{document}